\documentclass[aip,reprint,twocolumn]{revtex4-1}

\draft 

\usepackage{graphicx}
\usepackage{amsmath}
\usepackage[T1]{fontenc}

\newcommand{\beq}{\begin{equation}}
\newcommand{\eeq}{\end{equation}}
\newcommand{\beqa}{\begin{eqnarray}}
\newcommand{\eeqa}{\end{eqnarray}}
\newcommand{\vep}{\varepsilon}

\begin{document}


\title{Pauli Spin Blockade in a Resonant Triple Quantum Dot Molecule} 



\author{Yun-Pil Shim}
\email[Author to whom correspondence should be addressed: ]{yshim@utep.edu}
\affiliation{Department of Physics, The University of Texas at El Paso, El Paso, TX 79968, USA}


\date{\today}

\begin{abstract}
A Pauli spin blockade in quantum dot systems occurs when the charge transport is allowed only for some spin states, and it has been an efficient tool in spin-based qubit devices in semiconductors. We theoretically investigate a Pauli spin blockade in a triple quantum dot molecule consisting of three identical quantum dots in a semiconductor in the presence of an external magnetic field through the molecule. When the three-electron state is on resonance with two- or four-electron states, the Aharonov-Bohm oscillation and the Zeeman splitting lead to a periodic spin blockade effect. We focus on the spin blockade at a two- and three-electron resonance and show that we can tune the magnetic field to selectively allow only either a spin-singlet or spin-triplet state to add an additional electron from tunnel-coupled leads. This spin blockade maintains the three quantum dots at the optimal sweet spot against the charge noise, demonstrating its potential as an efficient readout scheme for the qubits in quantum dot systems. 
\end{abstract}


\maketitle 



%
%

%

\section{Introduction}

The controllability over the charge distribution and spins of electrons in quantum dot (QD) systems by electrical means \cite{jacak_qd1998,kouwenhoven_rpp2001,hanson_rmp2007,hsieh_rpp2012} offers great opportunity to study such small systems where a strong electron-electron interaction and quantum mechanical effects play important roles. As a prominent example, localized spins in QDs are one of the leading candidates for quantum computing devices \cite{loss_divincenzo_pra1998,kane_nature1998} and have been progressing rapidly.\cite{spin_qubit_review,Kloeffel_Loss_annurevcmp2013,burkard_arxiv2021}

Spin qubit devices rely on the spin-to-charge conversion and the charge detecting for qubit readouts.\cite{elzerman_nature2004,barthel_prl2009} The Pauli spin blockade (PSB) in a double quantum dot (DQD) system \cite{ono_austing_science2002} is widely used for this purpose. The PSB occurs when the transport of electronic charge is blocked due to the spin properties of the system. In a DQD with two electrons, when the system is in the charge configuration (1,1) and the (0,2) is the ground state, an electron is energetically allowed to tunnel from the left dot to the right dot. Since tunneling typically conserves the spin of the system and the (0,2) state is a spin singlet, the (1,1) singlet state can transition into the (0,2) singlet state by tunneling of an electron, while the (1,1) triplet state is not allowed to transition into the (0,2) singlet state. This spin blockade was experimentally demonstrated in a transport measurement in both vertical and lateral DQD systems.\cite{ono_austing_science2002,johnson_petta_prb2005} Combined with a charge sensor, the PSB can be used to read out the spin state of the spin qubit.\cite{johnson_petta_prb2005}

In semiconductor qubit devices, the single-qubit gates of individual spin qubits are experimentally challenging because it requires local modulation of magnetic properties.\cite{pioro-ladriere_tarucha_nphy2014,tokura_tarucha_prl2006,takeda_tarucha_sciadv2016,kawakami_vandersypen_nnano2014} Instead of using individual spins as qubits, the qubit state can be encoded in a two-dimensional subspace of a larger system consisting of a few QDs. These encoded qubits, such as the singlet-triplet (ST) qubit, \cite{Petta2005,Maune_Hunter_nature2012,Shulman_Yacoby_science2012} the exchange-only qubit, \cite{divincenzo_bacon_nature2000,Medford_Marcus_nnano2013,Eng_Hunter_sciadv2015,HRL2022} the resonant-exchange (RX) qubit, \cite{RX_theory_Taylor_prl2013,RX_exp_Medford_prl2013,RX_2q_gate_Doherty_prl2013} the always-on-exchange-only (AEON) qubit, \cite{shim_tahan_prb2016,malinowski_kuemmeth_prb2017} allow for easier control of the qubit states and provide some protection against environmental noise. \cite{Duan_Guo_prl1997,Zanardi_Rasetti_prl1997,Duan_Guo_pra1998,Lidar_Chuang_Whaley_prl1998,bacon_whaley_prl2000,kempe_whaley_pra2001,kempe_whaley_qic2001} Although the standard PSB and charge sensing can be used to read out these encoded qubits, it may be beneficial to investigate different types of spin-charge correlations that can be used for more efficient readout.  

Triple quantum dot (TQD) systems provide natural scaling up from DQD systems, and recent experimental progress in QD spin qubit devices \cite{APL_kandel2021,PhysRevLett.126.017701,PhysRevX.10.031006,PhysRevX.11.041025,PhysRevB.102.155404,HRL2022,Delft2022} makes it a viable option to use multiple QD molecules as a platform for quantum applications. TQD molecules are a minimal system with nontrivial geometrical effects.
Earlier experiments on TQD systems \cite{gaudreau_studenikin_prl2006,ihn_sigrist_njphys2007,schroer_greentree_prb2007,Rogge_NJP2009,amaha_prb2012,Gaudreau_nphys2012} demonstrated that they can show nontrivial features, such as interference effects. TQD systems in  ringlike geometry, interference effects are expected to play an important role in determining the electronic and spin structures. \cite{delgado_shim_prb2007,hsieh_rpp2012} In a resonant TQD, which is defined as the case where the lowest energy levels of the three QDs are equal, two-electron and four-electron complexes show a periodic transition between spin-singlet and triplet ground states due to the Aharonov-Bohm (AB) oscillation \cite{aharonov_bohm_pr1959} of the states in the presence of a perpendicular magnetic field. This periodic spin transition is of great consequence in transport through the system and coherent transport through a TQD system was experimentally observed showing AB oscillations in current.\cite{gaudreau_sachrajda_icps2006,ihn_sigrist_njphys2007} When the system is half-filled with three electrons, AB oscillations are suppressed due to the large Coulomb repulsion and the three-electron complex exhibits one transition from total spin $1/2$ to $3/2$ due to the Zeeman energy. 



\begin{figure}
\includegraphics[width=\linewidth]{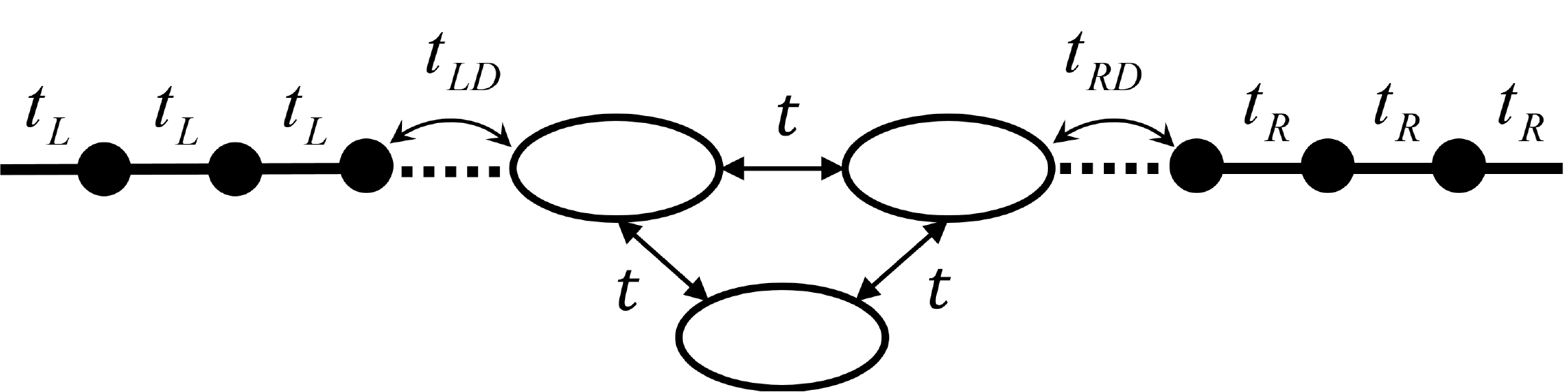}
\caption{ Schematic diagram of a triple quantum dot molecule connected to leads.
          $t$'s are the tunneling matrix elements that connect two different sites.}
\label{fig:system}
\end{figure}

In this paper, we study the effect on the transport of these different behaviors of two- and three-electron complexes in a spin transition. A resonant TQD molecule is connected to two semi-infinite chains of noninteracting leads (Fig.~\ref{fig:system}), with dot 1 connected to the left lead and dot 3 connected to the right lead. We study transport around a quadruple point (QP) where four charge configurations (1,1,0),(1,0,1),(0,1,1), and (1,1,1) in terms of the electron occupancy in each dot are all on resonance. A QP is defined as a point where four different charge configurations have the same probabilities. The periodic spin transition of the two-electron complex and the single spin transition of the three-electron complex lead to a periodic spin blockade in transport. When the two-electron state is in a spin singlet and the three-electron state is in spin $3/2$, the transport is suppressed because the addition of one electron cannot change the total spin from 0 to 3/2. Otherwise, the transport is allowed since all four charge configurations are on resonance. When the spin blockade occurs, we find triangular shape conductance islands around the QP in the Coulomb diamond at a very low temperature, and we explain this using the concepts of the transport channel and the trap state in multi-channel transport. \cite{shim_delgado_prb2009} We use a rate equation-based approach \cite{averin_korotkov_prb1991,beenakker_prb1991,bonet_deshmukh_prb2002,turek_matveev_prb2002,%
mitra_aleiner_prb2004,koch_oppen_prb2004,muralidharan_datta_prb2007} to calculate the transport by sequential tunneling between the leads and the TQD. A more detailed description of the method was given in our previous work. \cite{shim_delgado_prb2009} A master equation for the probabilities of each quantum many-body state of the system is solved at steady-state conditions.

PSB in TQDs have been studied previously, in a linear geometry, \cite{hsieh_prb2012,PhysRevB.89.161402,Busl2013_nnano,PhysRevLett.112.176803} in a detuned triangular TQD with two electrons \cite{PhysRevB.81.121306} or with three terminals. \cite{PhysRevB.96.155414} The PSB mechanism we present in this work is different from the previous work since it is due to the AB oscillations of the spin states in a resonant TQD. Maintaining the TQD on resonance allows us to use this effect for exchange-only qubits at sweet spots.

After briefly explaining our model Hamiltonian and the method of calculating the transport current in Sec. \ref{sec:Method}, we present the numerical results around the QP involving two- and three-electron states in Sec. \ref{sec:spin_blockade}. In Sec. \ref{sec:cond_island}, the triangular conductance island around the QP is presented and explained. A summary will be given in Sec. \ref{sec:Conclusion}.

\section{Method}\label{sec:Method}

We use a similar approach that was developed in our previous work \cite{shim_delgado_prb2009} to describe the triple dot molecule connected to the leads. A brief description is given here.

The resonant TQD is described by a Hubbard Hamiltonian with one orbital per dot,
\begin{eqnarray} \label{eq:Htqd} 
\widehat{H}_{TQD}
&=& \sum_{i=1}^{3}\sum_{\sigma} \left( \vep_0 + g^* \mu_B B \sigma \right)  d_{i\sigma}^{\dag} d_{i\sigma} \\
&+&  \sum_{i\neq j}\sum_{\sigma} t_{ij}(B) d_{i\sigma}^{\dag} d_{j\sigma}  
 + \sum_{i} U \hat{n}_{i\downarrow} \hat{n}_{i\uparrow}
+ {\frac{1}{2}} \sum_{i\neq j} V \hat{\rho}_{i} \hat{\rho}_{j}~,\nonumber
\end{eqnarray}
where the operators $d_{i\sigma}$ ($d_{i\sigma}^\dag$) annihilate (create)
an electron with spin $\sigma=\pm 1/2$ on orbital $i$ ($i=1,2,3$).
$\hat{n}_{i\sigma} = d_{i\sigma}^\dag d_{i\sigma}$
and $\hat{\rho}_{i} = \hat{n}_{i\downarrow} + \hat{n}_{i\uparrow}$
are, respectively, the spin and charge density on orbital level $i$.
$\vep_0$ is the energy level of individual QD at zero magnetic field,
which can be changed by applying proper voltages to external gates.
Notice that $\vep_0$ changes the $N$ electron energies
with respect to the $N+1$ energy levels.
$g^*$ is the effective Land\'e $g$-factor, and $\mu_B$ is the Bohr magneton.
$U$ is the on-site Coulomb repulsion and
$V$ is the direct Coulomb interaction between two electrons in different dots.
The hopping matrix elements are given by $t_{ij}(B)=t e^{2\pi i \phi_{ij}}$
where $\phi_{ij}$ is the Peierls phase factor. \cite{peierls_zphys1933,luttinger_pr1951}
We assume that the three quantum dots are located in the corners of an equilateral
triangle, and then, we have $\phi_{12}=\phi_{23}=\phi_{31}=-\phi/3$ and
$\phi_{ji}=-\phi_{ij}$, where $\phi_B=BA/\phi_0$ is the number of magnetic
flux quanta threading the system, with $A$ being the area of the triangle
and $\phi_0=h/e$ being the magnetic flux quantum.
The energy spectrum of the isolated TQD in the presence of the magnetic field
is obtained by the full configuration-interaction (CI) method. \cite{delgado_shim_prb2007}

The TQD system ($D$) is connected to leads ($r$=$L,R$) on both sides
with the left lead connected to dot 1 and the right lead to dot 3 (Fig. \ref{fig:system}).
Leads $r=L,R$ are described by non-interacting one-dimensional chains,
\begin{equation}
\nonumber
\widehat{H}_r = \sum_m\sum_{\sigma} \vep_0^r c^{\dag}_{m\sigma} c_{m\sigma}
               +\sum_m\sum_{\sigma} \left( t_r c^{\dag}_{m\sigma} c_{m+1\sigma} + h.c. \right)~,\label{eq:Hlead}
\end{equation}
where $m$ is from $-N_a$ to $-1$ for the left lead and from 1 to $N_a$ for the right lead.

The TQD and the leads are connected by tunneling Hamiltonian,
\beqa
\widehat{H}_{rD} &=& \sum_{\sigma} \left( t_{rD} c_{m_0\sigma}^{\dag} d_{i_0\sigma} + h.c \right)~,\label{eq:Hcouple}
\eeqa
where $m_0$ ($-1$ for $r=L$ and $1$ for $r=R$) and $i_0$ (1 for $r=L$ and 3 for $r=R$)
are the two adjacent sites of the lead and the dot connected by the tunneling
and $t_{rD}$ is the tunneling element connecting the two sites.
We will assume that tunnel coupling $t_{rD}$ is small enough that we can use a sequential tunneling picture
described in the following.

The current of electrons with spin $\sigma$ through the TQD is given by
\beqa
I^{\sigma}_{r \rightarrow D}
&=& (-e) \left( \Omega_{r\rightarrow D}^{\sigma} - \Omega_{D\rightarrow r}^{\sigma} \right)~,
\eeqa
\begin{widetext}
where
\beqa\Omega_{r\rightarrow D}^{\sigma}
&=& \sum_{N=0}^{5} \sum_{\alpha_N} \sum_{\alpha_{N+1}}
    f_r\left(E_{\alpha_{N+1}}^D-E_{\alpha_{N}}^D\right) P_{\alpha_N}
    \Gamma_r^{\sigma}\left(\alpha_N,\alpha_{N+1}\right) \\
\Omega_{D\rightarrow r}^{\sigma}
&=& \sum_{N=0}^{5} \sum_{\alpha_N} \sum{\alpha_{N+1}}
    \left[1-f_r\left(E_{\alpha_{N+1}}^D-E_{\alpha_{N}}^D\right)\right] P_{\alpha_N+1}
    \Gamma_r^{\sigma}\left(\alpha_N,\alpha_{N+1}\right)~,
\eeqa
and $f_r(\vep)=1/\{\exp[ (\vep-\mu_r)/k_BT ] +1 \}$ is the Fermi-Dirac distribution function
with respect to the chemical potential $\mu_r$ of lead $r$
and $P_{\alpha_N}$ is the probability of an $N$-electron state $\alpha_N$.
The coupling strength $\Gamma_r^{\sigma}$ is
\beqa
\Gamma_r^{\sigma}\left(\alpha_N,\alpha_{N+1}\right)
&=& \frac{2\pi |t_{rD}|^2}{\hbar|t_r|}
    \left| \langle \alpha_{N+1} | d^{\dag}_{i_0\sigma} | \alpha_N \rangle
    \right|^2
    \rho_{r\sigma}\left(E_{\alpha_{N+1}}^D -  E_{\alpha_{N}}^D\right)~.
\eeqa
where $\rho_{r\sigma}$ is the density of states of spin $\sigma$ per site in lead $r$.
The probabilities of each many-body states of the TQD molecule are obtained by solving the master equation.
The time evolution of the probabilities $P_{\alpha_N}$ is given by the following set of master equations:
\begin{eqnarray}\label{eq:Master}
\frac{dP_{\alpha_{N}}}{dt}
&=&\; \sum_{\alpha_{N+1}} \sum_{r=L,R} P_{\alpha_{N+1}}\left[ 1-f_r\left(E_{\alpha_{N+1}}^D -  E_{\alpha_{N}}^D\right) \right]
                          \Gamma_r(\alpha_{N},\alpha_{N+1}) 
     -\sum_{\alpha_{N+1}} \sum_{r=L,R} P_{\alpha_{N}} f_r\left(E_{\alpha_{N+1}}^D -  E_{\alpha_{N}}^D\right)
                          \Gamma_{r}(\alpha_N,\alpha_{N+1}) \nonumber\\
&+&   \sum_{\alpha_{N-1}} \sum_{r=L,R} P_{\alpha_{N-1}} f_r\left(E_{\alpha_{N}}^D -  E_{\alpha_{N-1}}^D\right)
                          \Gamma_{r}(\alpha_{N-1},\alpha_{N}) 
     -\sum_{\alpha_{N-1}} \sum_{r=L,R} P_{\alpha_{N}}\left[ 1-f_r\left(E_{\alpha_{N}}^D -  E_{\alpha_{N-1}}^D\right) \right]
                          \Gamma_r(\alpha_{N-1},\alpha_{N})~, \nonumber
\end{eqnarray}
\end{widetext}
where $\Gamma_r = \sum_{\sigma}\Gamma_r^{\sigma}$.
The initial probabilities at $t=0$ are given by equilibrium values,
\beqa
&& P_{\alpha_N}(0) = P_{\alpha_N}^{\rm eq}
 = \frac{ \exp\left( -\frac{ E_{\alpha_N}^D - \mu_0 N }{k_B T} \right)}{Z}~,\label{eq:P_eq}
\eeqa
where $Z$ is the grand partition function.
We find the probabilities at steady state $(t\rightarrow \infty)$ where all the time derivatives are zero, and then calculate the steady-state current using them.

\section{Spin blockade in a resonant TQD}\label{sec:spin_blockade}

To be specific, we study a resonant TQD in GaAs. 
We use the effective Rydberg $Ry$, which is $Ry=5.93$ meV for GaAs as the unit of energy.
The three QDs are located at the vertices of an equilateral triangle
with the distance between any two dots being 100 nm.
The magnetic field is measured with the number of flux quanta $\phi_B$ through the triangle.
We use parameters $U=2.5$, $V=0.5$, $t=-0.05$, and $t_L=t_R=-4.0$.
We assume that $t_{LD}=t_{RD}$, and they are small enough to justify the sequential tunneling picture.
The current is calculated in the unit of $I_0=e|t_{LD}|^2/\hbar |t_L|$.
In this effective unit, the current does not depend on the dot-lead tunneling $t_{LD}$ and $t_{RD}$.
The effective $g$-factor is taken to be $-0.44$ for GaAs.
The chemical potentials of each leads are
$\mu_L = \mu_0 + eV_{\rm sd}/2$ and $\mu_R = \mu_0 - eV_{\rm sd}/2$,
and we set the equilibrium chemical potential $\mu_0$ to be zero.
All the calculations are done at temperature 50mK, unless specified otherwise.


\begin{figure}
  \includegraphics[width=\linewidth]{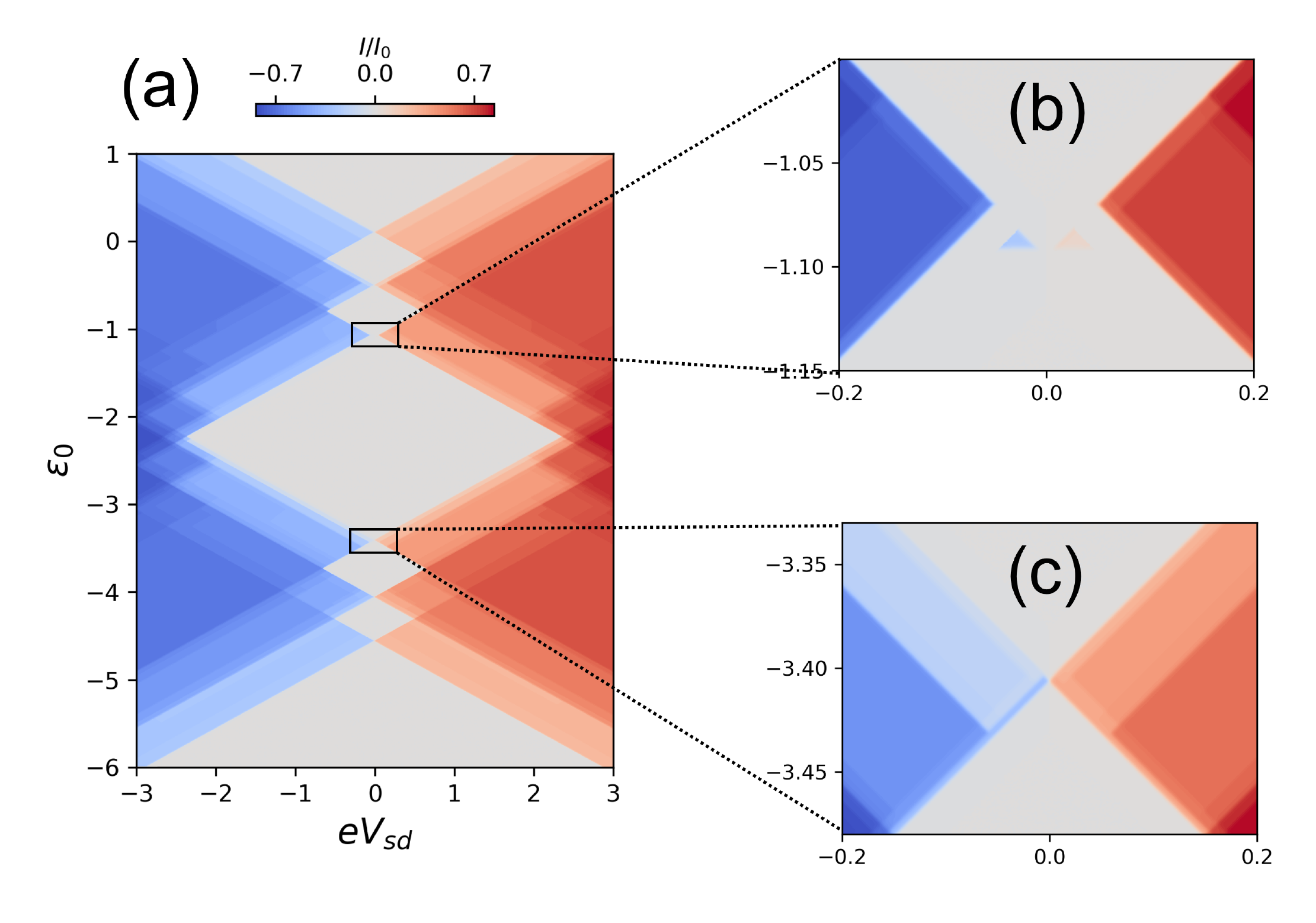}\\
  \caption{Current as a function of the bias $V_{\rm sd}$ and the overall shift $\vep_0$.
           Magnetic field is $\phi_B$=3.0.
           (a) Coulomb diamonds where no current flows due to the Coulomb blockade can be clearly seen.
           The largest diamond corresponds to $N=3$, and the addition energy is the largest for adding the fourth electron
           due to the large on-site Coulomb repulsion $U$.
           (b) Near the Coulomb blockade peak for adding a third electron.
           This peak is suppressed by the spin blockade
           since a two-electron state is a spin singlet and a three-electron state has $S_z=3/2$.
           (c) Near the Coulomb blockade peak for adding a fourth electron.
           The four-electron state is in a spin triplet, and there is no spin blockade effect here.}
  \label{fig:RTQD_Diamond}
\end{figure}

Figure \ref{fig:RTQD_Diamond} shows the current across the source and the drain 
as we change the overall energy shift $\vep_0$ and the bias $V_{\rm sd}$ between the leads
at magnetic field $\phi_B$=3.
At this magnetic field, the two-electron complex ground state is a spin singlet,
and a four-electron complex has a spin-triplet ground state. \cite{delgado_shim_prb2007}
We can observe the Coulomb diamonds [Fig.~\ref{fig:RTQD_Diamond}(a)] for this TQD molecule.
The open area on the top (bottom) corresponds to zero (six) electron in the molecule
and the diamonds correspond to electron numbers 1--5 in the molecule.
The large diamond for $N=3$ corresponds to the large addition energy
due to the strong on-site Coulomb repulsion $U$.
As we change the overall energy shift $\vep_0$,
the relative energy differences between $N$ electron and $N+1$ electron states change accordingly.
If $\vep_0$ increases (decreases),
the energies of $N+1$ electron states move up (down) with respect to $N$ electron states.
At fixed bias $V_{\rm sd}$, the relative change of the energy difference
leads to the creation of new transport channels and the destruction of some old transport channels.
This change in transport channels results in the change in current.
At fixed $\vep_0$ inside a Coulomb diamond, we need a finite bias for transport
to overcome the Coulomb blockade.

Interesting features appear near the Coulomb blockade peak for adding the third electron at around $\vep_0=-1.10$.
The four charge configurations that are degenerate at this quadruple point are (1,1,0),(1,0,1),(0,1,1), and (1,1,1).
The Coulomb blockade peak at this QP is suppressed, and we need a finite bias for transport.
This is due to the spin structure of the system and
can be observed in the zoomed-in picture[Fig.~\ref{fig:RTQD_Diamond}(b)].
Since the tunneling Hamiltonian equation.~\eqref{eq:Hcouple} conserves the $z$ component of the electron spin,
a transport channel involving an $N$ electron state and an $N+1$ electron state
is forbidden whenever the difference in the $z$ component of the total spin between these two states
is larger than 1/2.
At this QP, the total spin of the ground state with $N=2$ is 0, and the total spin of the $N=3$ ground state is $S=3/2$ and $S_z=3/2$:
therefore, the transition between these two states by adding an electron from the lead is forbidden.
On the other hand, when we add a fourth electron, the ground state of the four-electron complex at $\phi_B=3$ is a spin triplet,
and the transition is allowed. Therefore, there is no spin blockade[Fig.~\ref{fig:RTQD_Diamond}(c)].
If the bias voltage increases, more states are within the transport window, and the spin blockade will be lifted. 
A detailed analysis of the transport channels at a finite bias is given in Sec.~\ref{sec:cond_island}.


\begin{figure}
  \includegraphics[width=\linewidth]{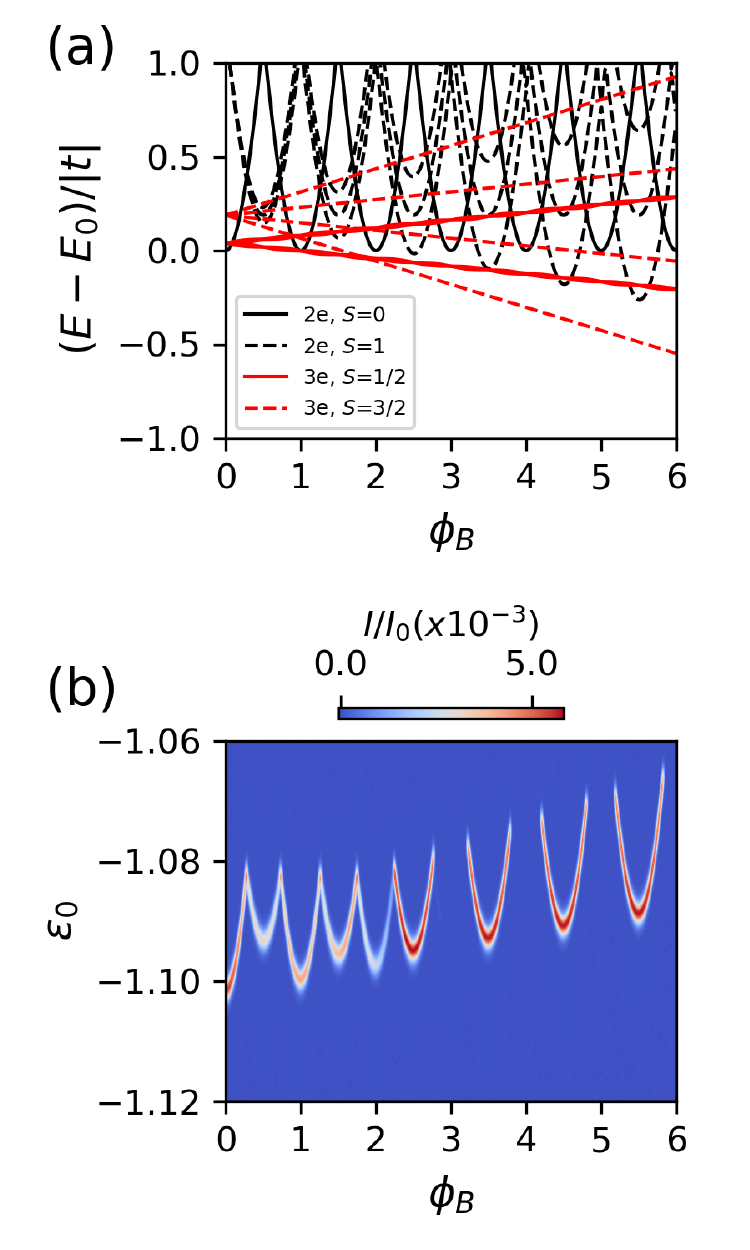}\\
  \caption{ Energy spectrum and linear current at spin blockade.
            (a)Black curves are the energy levels of two-electron states and red curves are for three-electron states.
            The two-electron ground state shows AB oscillations and repeated singlet-triplet transitions,
            while there is only one spin transition from $S=1/2$ to $S=3/2$ for the three-electron ground state.
            (b) Current as a function of the overall shift $\vep_0$ and the number of magnetic flux quanta $\phi_B$
            in the linear response regime ($eV_{\rm sd}=1.0\times 10^{-4}$).
            The current flows whenever two-electron ground state and three-electron ground state are on resonance
            and the difference in total spin is less than 1. Spin blockade occurs around integer $\phi_B=3,4,5,6$.
            At around $\phi_B=2$, current flows because $S=1/2$ state is in the thermal window for transport.}
  \label{fig:RTQD_linear}
\end{figure}

The ground state of the two- and four-electron complex alternates
between a spin singlet and a spin triplet due to the AB oscillations.
Figure \ref{fig:RTQD_linear}(a) shows the energy spectrum at the QP
corresponding to adding a third electron.
The overall shift $\vep_0=-1.10$ at this QP.
The two electron states (black curves) show AB oscillations,
and ground state alternates a spin singlet and a spin triplet.
On the other hand, the three-electron states (red curves) show
a spin transition from $S=1/2$ to $S=3/2$ due to the Zeeman energy.
Therefore, for a large enough magnetic field where the three-electron ground state has $S_z=3/2$,
the spin blockade occurs whenever the two-electron ground state is a spin singlet.
In the linear response regime, only the transport channel of the two-electron ground states
and the three-electron ground state is accessible.
The current shows clear spin-blockade effects as shown in Fig.~\ref{fig:RTQD_linear}(b).
It shows the current as a function of the overall shift $\vep_0$ and the magnetic field.
The current flows when the ground states of two- and three-electron cases are resonant.
At low magnetic fields, the $N=3$ ground state has $S=1/2$ and there is no spin blockade.
Once the Zeeman energy makes $S=3/2$ the ground state for $N=3$, the linear current is suppressed
around each integer flux quanta where the two-electron ground state is a spin singlet.
Similar behavior is observed when we add a fourth electron since the four-electron ground state
also alternates a spin singlet and a spin triplet.
At magnetic fields corresponding to (half) integer flux quanta,
the two-electron (four-electron) ground state is a spin singlet (triplet)
and the four electron ground state is a spin triplet (singlet).
This spin blockade leads to the suppression
of the third (fourth) electron Coulomb blockade peak at (half) integer flux quanta.

This PSB depends on the size of the TQD molecule since the Zeeman splitting between different total spin states does not depend on the geometry of the TQD molecule, while the magnetic flux through the TQD depends on the size of the molecule. Therefore, we can make the TQD molecule of a right size to tune the magnetic field where the PSB occurs in the system.

\section{Conductance island around a spin-blockade point at low temperature}\label{sec:cond_island}


\begin{figure}
  \includegraphics[width=\linewidth]{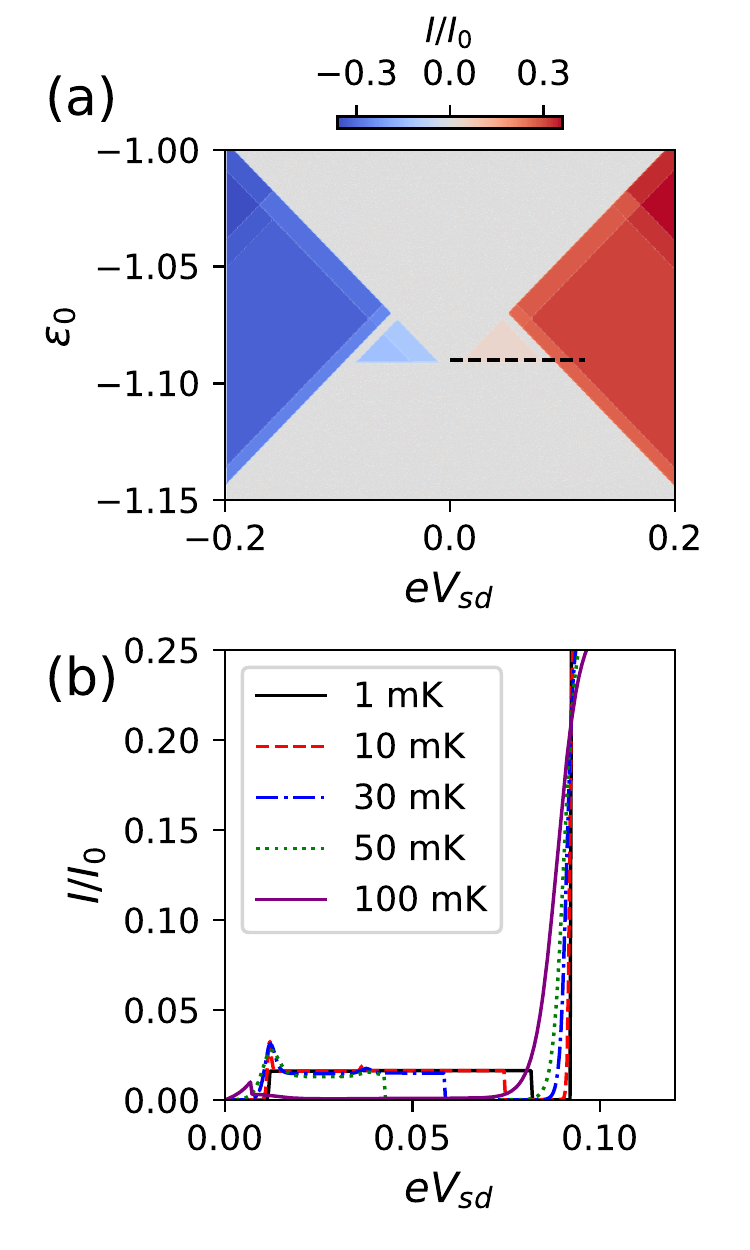}\\
  \caption{(a) Current as a function of the bias $V_{\rm sd}$ and the overall shift $\vep_0$ 
           near the spin-blockade region near zero temperature($T$=1.0 mK).
           There are triangular regions on nonzero current,
           which is suppressed at higher temperatures [Fig.~\ref{fig:RTQD_Diamond}(b)].
           (b) Current along the dotted line in (a) at different temperatures.
           We can see that the triangular region shrinks as temperature increases and finally disappears. }
  \label{fig:RTQD_spin_blockade_Diamond_T}
\end{figure}

Now, let us look closely at the spin-blockade region in the Coulomb diamond.
At a very low temperature ($T = 1$ mK), the spin blockade shows nontrivial features
[Fig.~\ref{fig:RTQD_spin_blockade_Diamond_T}(a)].
There are two small triangular areas: one in the forward bias region and the other one in the backward bias region,
with nonzero current.
Since we assume that $t_{LD}=t_{RD}$, there is a symmetry between forward and backward biases,
and we will explain it only for the forward bias case.
Figure~\ref{fig:RTQD_spin_blockade_Diamond_T}(b) shows the current
along the dotted line in Fig.~\ref{fig:RTQD_spin_blockade_Diamond_T}(a)
at different temperatures.
Rather counter-intuitively, the conductance triangle is present only at very low temperatures,
and it disappears as the temperature increases.


\begin{figure}
  \includegraphics[width=\linewidth]{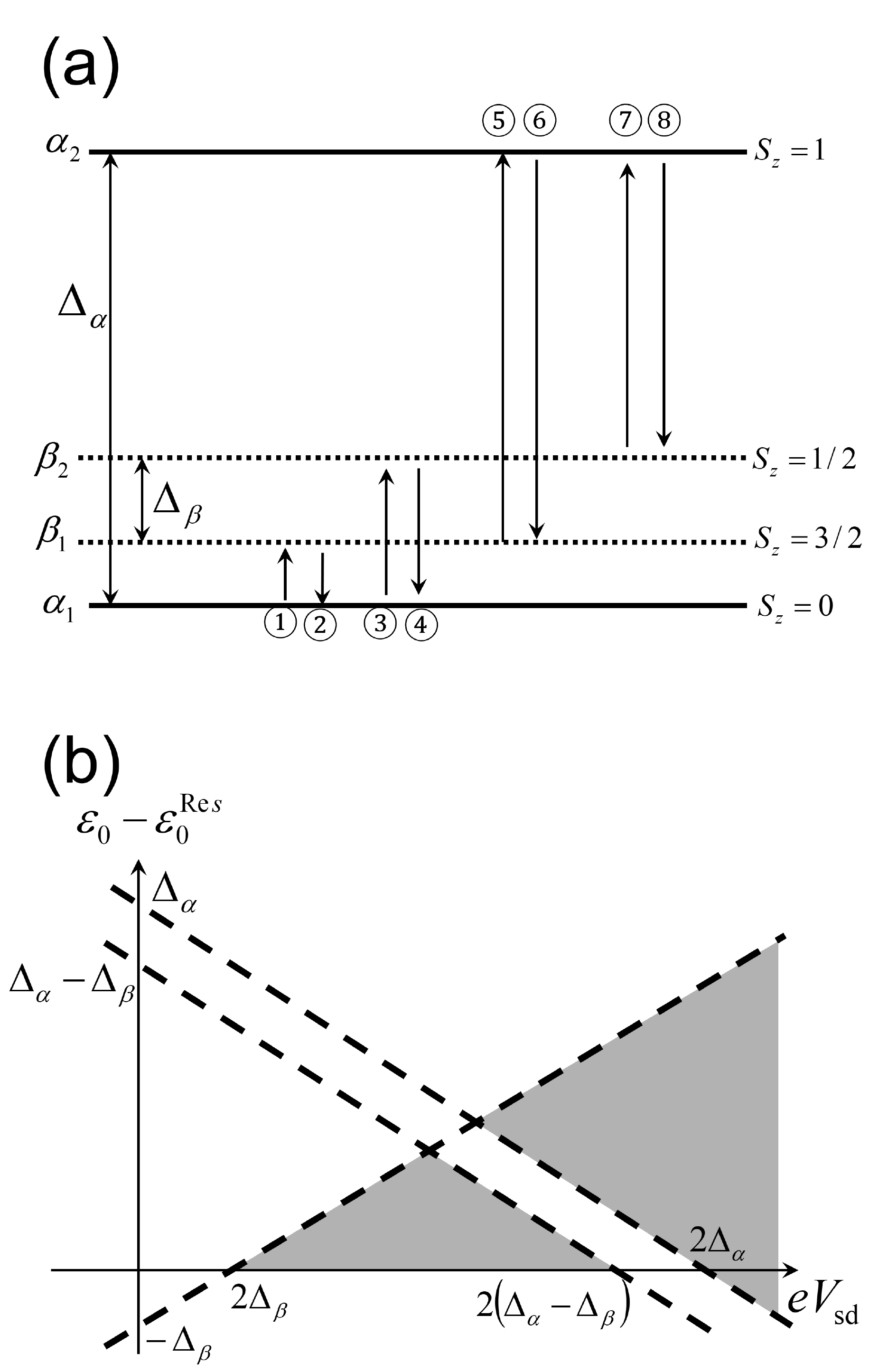}\\
  \caption{(a) Schematic diagram of energy levels near the spin-blockade region.
            Transitions \textcircled{1} and \textcircled{2} are spin-blockaded,
            and all the other transitions are allowed at a large enough bias.
            When $\alpha_2$ is an active state and the transition \textcircled{5} is not allowed, $\beta_1$ becomes a trap state
            and the current does not flow.
            (b) Transport regions using the conditions described in Sec. \ref{sec:cond_island}.
            The triangular region is separated from the large bias region
            due to the possibility that $\beta_1$ can be a trap state.}
  \label{fig:RTQD_spin_blockade}
\end{figure}

The conductance island can be explained using the transport conditions described in Ref.~\cite{shim_delgado_prb2009}.
Transport can be understood as sequential tunneling events: an electron tunnels from a lead to the TQD in an $N$-electron state $\alpha_N$, transitioning the TQD state into an $(N+1)$-electron state $\beta_{N+1}$, and then an electron tunnels from the TQD into the other lead and the TQD returns to its original state $\alpha_N$. We define this a transport channel $(\alpha_N,\beta_{N+1})$.
An active transport channel $(\alpha_N,\beta_{N+1})$ should satisfy
\beqa
&& \mathrm{(i)} \quad \left| E_{\beta_{N+1}}-E_{\alpha_N} -\mu_0 \right| \leq \frac{eV_{\rm sd}}{2}~, \label{eq:transport_window}\nonumber\\
&& \mathrm{(ii)} \quad \Gamma_L(\alpha_N, \beta_{N+1}) \neq 0
        ~ \mathrm{and} ~ \Gamma_R(\alpha_N, \beta_{N+1}) \neq 0, ~ \mathrm{and}\label{eq:nonzero_transition_rate}\nonumber\\
&& \mathrm{(iii)} \quad \text{$\alpha_N$ or $\beta_{N+1}$ can be populated.}\nonumber
\eeqa
Any state that participates in one or more transport channels will be called an {\it active state}.
Condition (i) specifies that the transition between $\alpha_N$ and $\beta_{N+1}$ by adding or subtracting an electron from the source or drain is energetically allowed. Condition (ii) allows the tunneling of an electron from and to the source/drain. Condition (iii) is required so that the two states can participate actively to the transport. The system is in the ground state initially. Other states can be populated either thermally or by transition through the addition/subtraction of an electron from the leads. The tunnel coupling is generally allowed: therefore, we will consider conditions (i) and (iii) to determine transport conditions.

For the transport to occur, there must be one or more transport channels
and there must not be any trap state.
A state $\gamma_N$ is defined to be a {\it trap state} if
(i) the transition $\alpha_{N\pm 1} \rightarrow \gamma_N$ is possible for an active state $\alpha_{N\pm 1}$
by adding or subtracting an electron, but
(ii) the transition $\gamma_N \rightarrow \alpha_{N\pm 1}$ is not allowed for any active state $\alpha_{N\pm 1}$.

Let us first consider the temperature $T$=0 case. 
Near the resonant point ($\vep_0=\vep_0^{Res}$)
where the two-electron ground state and the three-electron ground state are on resonance,
we have four states that can participate in the transport [see Fig.~\ref{fig:RTQD_spin_blockade}(a)].
The spin singlet ($\alpha_1$) and triplet ($\alpha_2$) for two-electron systems
and a spin polarized state with $S_z=3/2$ ($\beta_1$) and an unpolarized state with $S_z=1/2$ ($\beta_2$)
can form various transport channels except for
($\alpha_1,\beta_1$), which is spin blockaded (transitions \textcircled{1} and \textcircled{2} are not allowed).
All the other transitions (\textcircled{3} $\sim$ \textcircled{8}) are allowed if the bias is large enough.
For an electron to transport, we need at least one transport channel
and there must be no trap state.
Therefore, $(\alpha_1,\beta_2)$ must be a transport channel,
and $\beta_1$ must not be a trap state.
First, let us determine the condition that $(\alpha_1,\beta_2)$ is a transport channel.
Condition (i) for the transport channel states
\beq
\left| E_{\beta_2}-E_{\alpha_1} \right| \le \frac{eV_{\rm sd}}{2} ~.\label{eq:a1b2}
\eeq
With $\vep_0 \ge \vep_0^{Res}$ ( i.e., $E_{\alpha_1} \le E_{\beta_1}$),
this is the only condition required for $(\alpha_1,\beta_2)$ to be a transport channel
since $\alpha_1$ is the ground state [($\alpha_1,\beta_1$) channel is spin blockaded and cannot form a transport channel].
For $\vep_0 \le \vep_0^{Res}$ ( i.e., $E_{\alpha_1} \ge E_{\beta_1}$), $\beta_1$ is the ground state.
In addition to Eq.~\eqref{eq:a1b2}, we need a mechanism to populate either $\alpha_1$ or $\beta_2$.
Since $\beta_1$ is the ground state and $\alpha_1$ is not allowed to form a transport channel with $\beta_1$,
we need $\alpha_2$ state to form a transport channel with $\beta_1$.
Then, $\alpha_2$ and $\beta_2$ will also form a transport channel.
This will satisfy condition (iii) for the transport channel $(\alpha_1,\beta_2)$.
Hence, we have the condition
\beq
\left| E_{\alpha_2}-E_{\beta_1} \right| \le \frac{eV_{\rm sd}}{2}
\quad \text{for}\; \vep_0 \le \vep_0^{Res}~.\label{eq:a2b1}
\eeq
Eqs.~\eqref{eq:a1b2} and \eqref{eq:a2b1} determine the transport region unless $\beta_1$ is a trap state.
In terms of $\Delta\vep_0 \equiv \vep_0-\vep_0^{Res}$ and $eV_{\rm sd}$ they become
\beqa
-\frac{eV_{\rm sd}}{2} - \Delta_{\beta} \le \Delta\vep_0 \le \frac{eV_{\rm sd}}{2} - \Delta_{\beta} \label{eq:transport1} \\
\Delta\vep_0 \ge -\frac{eV_{\rm sd}}{2} + \Delta_{\alpha} \quad \text{for}\; \Delta\vep_0 \le 0.\label{eq:transport2}
\eeqa
We now determine the conditions that $\beta_1$ is a trap state, which will forbid current flow
inside the transport region defined above.
The transition between $\beta_1$ and $\alpha_1$ is not allowed (Pauli spin blockaded). 
Therefore, $\beta_1$ is a trap state if $\alpha_2$ is an active state
and transition \textcircled{6} is allowed (to populate $\beta_1$ state), but transition \textcircled{5} is not allowed (to forbid further transition).
$\alpha_2$ is an active state if $(\alpha_2,\beta_2)$ is a transport channel, which requires
\beq
\left| E_{\beta_2}-E_{\alpha_2}\right| \le \frac{eV_{\rm sd}}{2}~.\label{eq:trap1}
\eeq
Transition \textcircled{6} is always allowed with a forward bias, and transition \textcircled{5} is blocked if
\beq
E_{\alpha_2}-E_{\beta_1} \ge \frac{eV_{\rm sd}}{2}.\label{eq:trap2}
\eeq
Eqs.~\eqref{eq:trap1} and \eqref{eq:trap2} can be expressed in terms of $\Delta\vep_0$ as
\beq
-\frac{eV_{\rm sd}}{2} + \Delta_{\alpha} - \Delta_{\beta}
\le \Delta\vep_0 \le -\frac{eV_{\rm sd}}{2} + \Delta_{\alpha}.\label{eq:trap}
\eeq
Inside the transport region defined by Eqs.~\eqref{eq:transport1} and \eqref{eq:transport2},
current will be suppressed in the region defined by Eq. \eqref{eq:trap} where $\beta_1$ is a trap state.
These conditions are plotted in Fig.~\ref{fig:RTQD_spin_blockade}(b), which
explains the result in Fig.~\ref{fig:RTQD_spin_blockade_Diamond_T}(a).
This trap state can be considered a generalization of the spin blockade state and analogous to the dark state in optical systems as was studied in the earlier work  \cite{Michaelis_2006,emary_prb2007,poltl_prb2009} for transport through TQD systems in different setups than presented here. It blocks or prevents further transition of the system.

As temperature increases, $\beta_1$ becomes a trap state more easily due to the thermal effects ({\it i.e.}, it can be thermally populated), and the conductance triangle shrinks and finally disappears
as was seen in Fig.~\ref{fig:RTQD_spin_blockade_Diamond_T}(b).
This thermal effect may seem counter-intuitive, but it depends on the energies of each levels whether the thermal effects lift the PSB or enhance it. In the current case, the thermal effect allows the trap state to be populated and induce the spin blockade.

\section{Conclusion}\label{sec:Conclusion}

We presented spin-blockade effects in a resonant TQD molecule and a triangular conductance island around the spin-blockade point. The PSB in a TQD is due to the fact that adding a single electron cannot change the total spin more than 1/2, which is qualitatively different from the standard PSB in DQD systems. The conductance island appearing at a very low temperature can be explained using the concept of a trap state. We considered only the GaAs-based QDs to be specific, but this TQD PSB should also be applicable to other materials, such as silicon QDs and hole spin qubits in germanium.

Depending on the spin state of the system, the TQD PSB leads to a different charge configuration with a different number of charges, which may be more easily detected by a charge sensor, similar to the charge latching readout mechanism. \cite{harvey_collard_prx2018,seedhouse_prxq2021} Unlike the DQD PSB, the TQD PSB presented here does not require detuning the QDs. The three QDs remain resonant allowing the system to remain on the natural sweet spot against charge noise. \cite{DQD_sweet_spot_Reed_Hunter_prl2016,DQD_symmetric_operation_Martins_Kuemmeth_prl2016,shim_tahan_prb2016,shim_prb2018}

%

\begin{acknowledgments}
The author thanks P. Hawrylak for helpful discussions.
\end{acknowledgments}

\bibliography{tqd_psb_arxiv_v2}

\end{document}